\documentclass[
  twocolumn,
  prb,
  amsmath,
  amssymb,
  superscriptaddress
]{revtex4}
\usepackage[colorlinks]{hyperref}
\usepackage{xcolor}
\usepackage[utf8]{inputenc}
\usepackage{amsmath}
\usepackage{amsfonts}
\usepackage{amssymb}
\usepackage{graphicx}
\usepackage{wrapfig}
\usepackage{MnSymbol}
\usepackage{braket}
\usepackage{comment}
\renewcommand{\v}[1]{\textbf{#1}}
\newcommand{\tr}{\text{tr}}

\newcommand{\PRLsection}[1]{\textcolor{blue}{\textit{#1}}}

\begin{document}

\title{Interaction-induced velocity renormalization in magic angle twisted trilayer graphene}

\author{Laura Classen}
\affiliation{Condensed Matter Physics and Materials Science Division, Brookhaven National Laboratory, Upton, New York 11973, USA}
\affiliation{Max Planck Institute for Solid State Research, D-70569 Stuttgart, Germany}

\author{J.~H.~Pixley}
\affiliation{Department of Physics and Astronomy, Center for Materials Theory, Rutgers University, Piscataway, New Jersey 08854, USA}

\author{Elio~J.\ K\"onig}
\affiliation{Max Planck Institute for Solid State Research, D-70569 Stuttgart, Germany}

\begin{abstract}

Twistronics heterostructures provide a novel route to control the electronic single particle velocity and thereby to engineer strong effective interactions. Here we show that the reverse may also hold, i.e.~that these interactions strongly renormalize the band structure. We demonstrate this mechanism for mirror-symmetric magic angle twisted trilayer graphene at charge neutrality and in the vicinity of a phase transition which can be described by an Ising Gross-Neveu critical point corresponding, e.g., to the onset of valley Hall or Hall order. While the non-interacting model displays massless Dirac excitations with strongly different velocities, we show that interaction corrections make them equal in the infrared. 
However, the RG flow of the velocities and of the 
coupling to the critical bosonic mode is strongly non-monotonic and dominated by the vicinity of a repulsive fixed point. We predict experimental consequences of this theory for tunneling and transport experiments and discuss the expected behavior at other quantum critical points, including those corresponding to intervalley coherent ordering.
\end{abstract}

\maketitle

\begin{figure}
    \centering
    \includegraphics[scale=1]{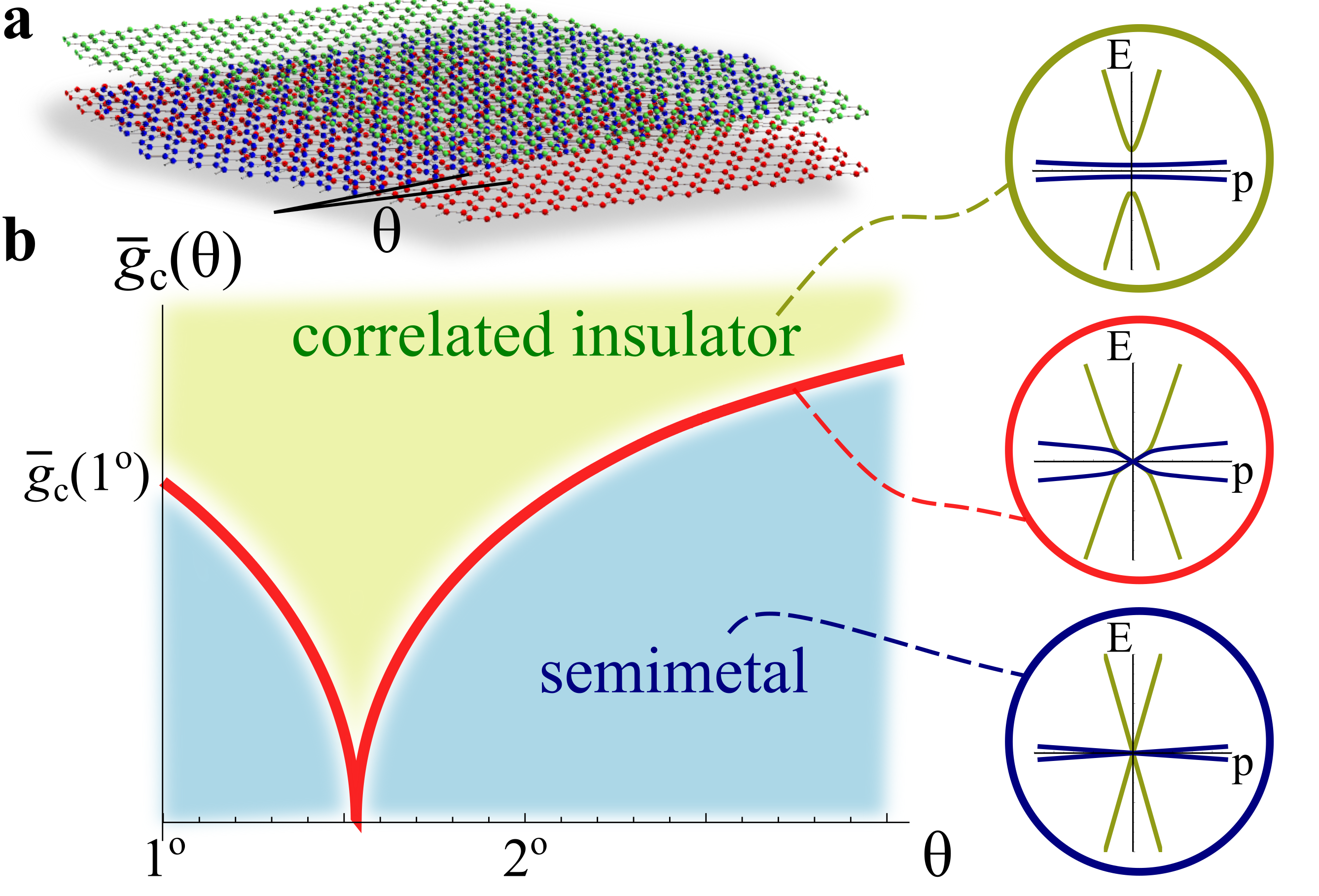}
    \caption{a) Schematics of symmetric TTG. b) Mean field phase diagram of the GN model, Eq.~\eqref{eq:L0}, as a function of twist angle $\theta$ for coupling strength $g_+ = g_- = \bar g$. Insets: Schematic dispersion relations in the semimetallic phase (blue), and in the gapped phase (correlated insulator, green). At criticality (red line), the dispersion is scale dependent and slow and fast fermions acquire the same, slow velocity in the infrared.}
    \label{fig:Schematics}
\end{figure}

\PRLsection{Introduction}:
Manipulating the nature of  low-energy electronic excitations in solid-state compounds has remained a long standing goal in condensed matter physics. Thanks to the recent advances in the isolation and manipulation of two-dimensional materials, it is now possible to dramatically modify the low energy band structure by twisting (i.e.~rotating) and stacking. 
As first demonstrated in twisted bilayer graphene (TBG), the moir\'e pattern formed from the interference between the twisted and stacked layers greatly renormalizes the excitation velocity\cite{Morell2010,BistritzerMacDoanld2011} allowing a small bare interaction scale to become dominant, which drives the formation of correlated insulating states and superconductivity\cite{CaoJarillo2018,CaoJarillo2018b}. This `twistronic' paradigm has now been extended well beyond graphene to, e.g., emulating Hubbard models using transition metal dichalgonides\cite{WangDean2020,Tang2020,PhysRevLett.121.026402,zhou2021quantum,scherer2021N4,PhysRevB.100.035413} or manipulating the superconducting state using cuprates~\cite{ZhaoKim2021,VolkovPixley2021,TummuruFranz2021}. 

Beyond  stacking and twisting two sheets of atoms, recent experiments have shown it is possible to twist various arrangements of a number of graphene layers~\cite{CaoJarillo2020,ShenZhang2020,LiuKim2020,park2021magicangle,ZhangNadjPerge2021} and also produce non-trivial phase diagrams with insulating and superconducting states. 
In particular, symmetrically twisted multi-layer graphene systems, where the layers are rotated by an alternating angle $\pm\theta$, can be considered as a sum of twisted bilayers.\cite{KhalafVishwanath2019} 
Thus, they offer an alternative setting where strongly renormalized narrow bands appear at a magic angle (where the Dirac velocity vanishes) as in TBG. For an odd number of layers, these TBG-like bands are supplemented by fast Dirac dispersing bands due to the symmetric nature of the system.\cite{KhalafVishwanath2019,CalaguruBernevig2021} For example, in symmetrically twisted trilayer graphene (TTG) the mirror eigenvalues in the setup of Fig.~\ref{fig:Schematics} a) protect the fast excitations from being renormalized by the twist. As a result, fast and slow Dirac excitations coexist in TTG near charge neutrality. 

Experiments on TTG have discovered a readily tunable superconducting ground state,
phases of broken symmetry at fillings of $\pm2$ electrons per moir\'e unit cell, 
and a sizeable resistance at charge neutrality.\cite{ParkJarillo2021,HaoKim2021} 
The spontaneous appearance of a gap at charge neutrality was not reported in TTG systems to date. 
But it was observed in some TBG devices without substrate alignment\cite{LuEfetov2019,Wu2021}, where it is believed to arise from interaction effects and expected based on the exact solution of effective theories\cite{LianBenervig2021}. The difference between insulating and semi-metallic behavior at charge neutrality may come from a strain-induced quantum phase transition. \cite{PhysRevLett.127.027601} 
Theoretically, different candidate states for a gap opening at charge neutrality in TBG have similar energies.  \cite{BultinckZalatel2021,BrillauxSavary2020,PhysRevX.11.011014,PhysRevB.102.035161,PhysRevB.98.081102,PhysRevB.102.035136,Liao_2021}
Among them, (sublattice-polarized) quantum valley Hall or intervalley coherent order have also been discussed to gap out the slow bands in TTG,
based on extensions of the effective, exactly soluble models from TBG to TTG\cite{ChristosScheurer2021}. 

When Dirac electrons reside close to a phase transition and interact with critical bosonic modes,
Lorentz symmetry is often emergent\cite{Roy2016}. This implies that different bare velocities must renormalize to become equal, and the average velocity may vanish~\cite{SitteGarst2009}, diverge~\cite{KoenigPixley2018}, or approach a finite value, as it occurs, for example, to the two Dirac velocities in spin 3/2 systems. 
\cite{RoyJuricic2018,Boettcher2020,Mandal2021} 
While in TTG devices, the vicinity to external gates implies effective short-range interactions, we highlight that 
velocity renormalization due to Coulomb interactions in suspended graphene was predicted theoretically\cite{SheehySchmalian2007,BarlasMacDonald2007,HwangDasSarma2007} and subsequently confirmed experimentally\cite{EliasGeim2011,Yu3282,ulybyshev2021bridging} more than a decade ago.
In contrast to the single-particle magic-angle phenomenon, many-body effects are crucial to the physics in the examples above. 
This raises the question how interactions affect the velocity renormalization in TTG. Do the velocities of fast and slow Dirac modes approach each other and does this work for or against the correlation effects in the TBG-like bands which are rooted in the vanishing Dirac velocity? 

In this paper we address the velocity renormalization combining the single-particle effects 
in a moir\'e system with the many-body physics of electron-electron interactions. 
To this end, we study an Ising Gross-Neveu (GN) theory with a generic ratio of fast and slow Dirac fermions employing a renormalization group (RG) procedure in $D=2+1$ dimensions. 
This describes the low energy properties of TTG near charge neutrality in the vicinity of a quantum critical point, which separates a Dirac semimetal phase from a gapped (valley) quantum Hall state. 
The critical interaction strength can be smoothly varied by changing the twist, which allows us to connect the untwisted critical point to the ground state at the magic-angle\cite{BrillauxSavary2020}, Fig.~\ref{fig:Schematics} b).
We demonstrate that slow and fast Dirac excitations cannot be considered separately as often done in TTG. As a result of their strong influence on each other the slow excitations become faster and the fast ones slower ensuring emergent Lorentz symmetry at the asymptotic critical point. 
Overall, the system becomes more correlated in the sense that the mean velocity decreases. More generally, the asymptotic 
mean velocity 
is determined by the ratio of the number of fast and slow modes. 
We estimate the velocity renormalization to be a measureable effect on length scales accessible in experiment and predict the simultaneous suppression of the quasiparticle weight.  

\PRLsection{Model.}
As an effective model for TTG near charge neutrality, we consider the generalized Gross-Neveu theory
($\hbar = 1$ throughout and $\sigma_{x,y,z}$ are Pauli matrices in sublattice space)
\begin{equation} \label{eq:L0}
\mathcal L = \sum_{\pm} \sum_{\alpha = 1}^{N_\pm} \bar \psi_{\alpha, \pm} [\partial_\tau + v_\pm \vec p \cdot \vec \sigma + g_\pm \phi \sigma_z] \psi_{\alpha, \pm} + \frac{N}{2} \phi^2,    
\end{equation}
which describes fast (+) and slow (-) Dirac fermions with $2N=(N_++N_-)$ flavors in total and different velocities $v_\pm$. To leading order, they are related to the twist angle via 
\begin{equation}
    \frac{v_-}{v_+} = \frac{1-3 \alpha^2}{1 + 6 \alpha^2}, \label{eq:BM}
\end{equation}
where $\alpha = \theta_c /\theta {\sqrt{3}}$.\cite{BistritzerMacDoanld2011, KhalafVishwanath2019} 
The magic angle defined by vanishing velocity occurs at $\theta = \theta_c \approx 1.6^\circ$. 
Interactions are mediated by the Ising order parameter field $\phi$, which is introduced via a Hubbard-Stratonovich transformation and corresponds to the interaction channel for a (sublattice polarized) quantum Hall or valley quantum Hall instability. 
By symmetry, $\phi$ couples simultaneously to both slow and fast fields, see Supplement~\cite{SuppMat}. 
The case $N_+ = 4$ (spin + valley) and $N_- = 8$ (spin + valley + mini-valley) applies to 
TTG in the vicinity of the magic angle. 
More generally, the model captures the dynamical mass generation in Dirac materials with an arbitrary number of slow and fast flavors near the onset of spontaneous order that breaks an Ising symmetry. 

\begin{figure}
    \centering
    \includegraphics{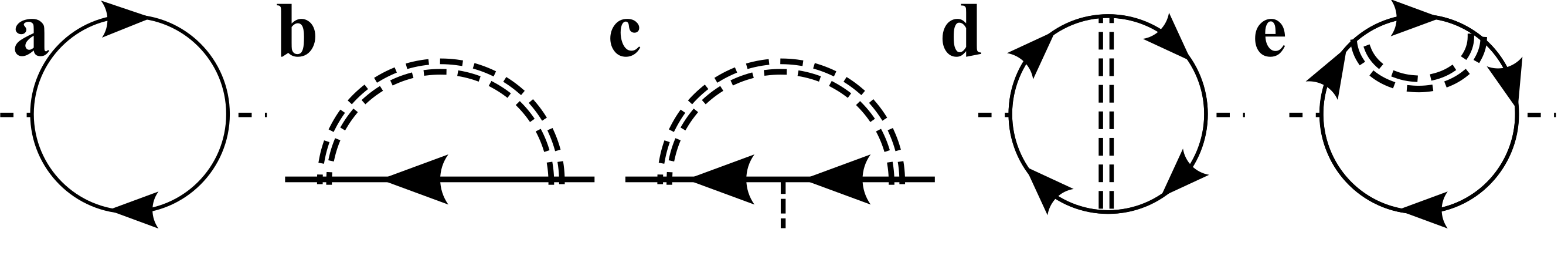}
    \caption{a) The polarization operator calculated with the non-interacting fermionic Green's function. It is $\mathcal O(N)$, while other diagrams are sub-leading at $N \rightarrow \infty$. This controls an RPA resummation, which is denoted by a double dashed line. b) Fermionic self-energy 
    and vertex correction 
    [$\mathcal O(1/N)$]. 
    d) Leading two-loop corrections to the bosonic self-energy [$\mathcal O(1)$].}
    \label{fig:Diagrams}
\end{figure}

\PRLsection{Mean field phase diagram.}
In the large $N$ limit, we can integrate out the fermions 
and calculate the effective bosonic mass in the saddle-point approximation 
\begin{equation}\label{eq:BosonicMassMainText}
    m = \frac{1}{g_+ g_-} - \sum_\pm \frac{N_\pm}{2 N \pi}\frac{g_\pm}{v_{\pm} g_\mp}  \Lambda, 
\end{equation}
where we use $\Lambda$ as a UV momentum cut-off scale. The second term in Eq.~\eqref{eq:BosonicMassMainText} stems from the fermionic susceptibility, Fig.~\ref{fig:Diagrams} a). A sign change in $m$ signals the dynamical mass generation and defines the critical coupling where the mean field phase transition occurs. 
Exploiting the relation in Eq.~\eqref{eq:BM} allows us to perturbatively determine the interaction driven transition as a function of twist angle and leads to the phase diagram plotted in Fig.~\ref{fig:Schematics} b). 
Importantly, this shows that at the magic-angle an infinitesimal interaction strength is sufficient to drive the mean field transition, a hallmark of the renormalized flat band that occurs in the slow Dirac excitations.
Right at the magic angle of TTG, the kinetic part of the slow sector is dominated by a quadratic band touching~\cite{Hejazi-2019} where the critical coupling vanishes. Therefore, logarithmic self-energy corrections introduce an additional sharpening\cite{RayJanssen2018} of the phase transition near $\theta = \theta_c$ (not shown).

\PRLsection{Quantum fluctuations.} 
We here summarize the strategy to incorporate quantum fluctuations near $m = 0$, 
in order to correctly capture the nature of the quantum critical point
(leaving technical details for the supplement\cite{SuppMat}).
This allows us to unveil an interaction driven renormalization of the velocities making them equal in the infared.
The contributions beyond mean field theory are organized in a systematic way via a
 large-$N$ expansion,  \cite{RosensteinPark1989,RosensteinPark1991,KhveshchenkoPaaske2001,doi:10.1142/S0217751X18300326,PhysRevB.99.195135} 
 which has the advantage of performing calculations directly in $D = 2 + 1$ and complements commonly used expansions near the upper critical dimension.\cite{RoyJuricic2018,Boettcher2020,Mandal2021}
As a starting point, it is exploited that the one-loop bosonic self energy (i.e. the susceptibility, Fig.~\ref{fig:Diagrams} a)) overpowers a quadratic kinetic term leading to an effective RPA resummed bosonic propagator to leading order in $N$
\begin{equation}
 D(\omega, \vec q) = \frac{1}{N}\left (m+\sum_{\pm}\alpha_\pm \sqrt{\omega^2 + v_\pm^2 \vec q^2} \right)^{-1},
 \end{equation} 
where $\alpha_\pm = N_\pm \rho_g^{\pm 1}\rho_v^{\mp 1}/N $ and we defined $\rho_g=g_+/g_-$ and $\rho_v=v_+/v_-$. The linear scaling of $D^{-1}$ leads to logarithmic divergencies in the next-to-leading-order diagrams in Fig.~\ref{fig:Diagrams} b) - e), which are formally cured by the inclusion of counter terms in the bare action. These counter terms renormalize all coupling constants as well as the fields in Eq.~\eqref{eq:L0} and determine the corresponding renormalization group (RG) equations. In addition to this field-theoretic RG procedure, we also performed Wilsonian momentum-shell-RG, yielding the same result.\cite{SuppMat}

\PRLsection{Renormalization group equations.} 
There are four running coupling constants, $v_{\pm}, g_\pm$, in addition to field renormalizations $Z_\psi^\pm, Z_\phi$, which introduce non-trivial scaling dimensions for three fields. 
It is convenient to express the RG equations in terms of mean and relative coupling constants $\rho_g = g_+/g_-, \rho_v = v_+/v_-, \bar v = \sqrt{v_+ v_-}$ and the bosonic gap $m$.
We find to leading order in the $1/N$ expansion~\cite{SuppMat}
\begin{subequations}
\begin{align}
    \frac{d \rho_g }{d \ln b} &= \rho_g\sum_\pm \frac{\pm 4}{3\pi^2 N}\left [ f_2^\pm-3f_1^\pm \right], \label{eq:BetaRhog}\\
    \frac{d \rho_v }{d \ln b} &=  \rho_v\sum_\pm \frac{\pm 2}{\pi^2 N}\left [f_2^\pm - f_1^\pm\right], \label{eq:BetaRhov}\\
    \frac{d \bar v }{d \ln b} &=  \bar v \sum_\pm \frac{1}{\pi^2 N} \left [f_2^\pm - f_1^\pm\right] \label{eq:Betav},\\
    \frac{d m}{d \ln b} &= \left \lbrace 1 -   \frac{8}{3\pi^2 N}  \left[3 f_3 - \sum_\pm \frac{3f_1^\pm - f_2^\pm}{2}\right] \right \rbrace m, \label{eq:Betam}
\end{align}
with $b = \Lambda/\Lambda '$ where $\Lambda'$ is the running scale. We have also absorbed a shift into the mass in Eq.~\eqref{eq:Betam} as compared to Eq.~\eqref{eq:BosonicMassMainText}. In addition, bosonic and fermionic fields are rescaled at one-loop level
\begin{align}
    \frac{d \ln Z_\psi^\pm}{d\ln b} &=- \frac{2}{3\pi^2 N} [3 f_1^\pm - 2 f_2^\pm], \label{eq:ZPsi}\\
    \frac{d \ln Z_\phi}{d\ln b} &=  \frac{4}{3\pi^2 N} \sum_\pm[3 f_1^\pm -  f_2^\pm]. \label{eq:ZPhi}
\end{align}
\label{eq:RGEquations}
\end{subequations}
In these equations, we have introduced the five functions $f_{1,2}^\pm =f_{1,2}^\pm(\rho_g, \rho_v,N_+/N_-), f_3= f_3(\rho_g, \rho_v,N_+/N_-)$, which are implicitly defined through the following integrals 
\begin{subequations}
\begin{align}
f_1^\pm &= \rho_g^{\pm 1} \int_{-\infty}^\infty dx \frac{1}{({x^2 + \rho_v^{\pm 1}})[\sum_\xi \alpha_\xi \sqrt{x^2 + \rho_v^{\xi}}]}, \\
f_2^\pm &= 3\rho_g^{\pm 1}  \int_{-\infty}^\infty dx \frac{\sum_{\xi = \pm} \alpha_\xi \frac{\rho_v^\xi/2}{\sqrt{x^2 + \rho_v^{\xi}}}}{({x^2 + \rho_v^{\pm 1}})[\sum_{\xi = \pm} \alpha_\xi \sqrt{x^2 + \rho_v^{\xi}}]^2},\\
f_3 &= \sum_\pm \int_{-\infty}^\infty dx \frac{\alpha_\pm}{\sqrt{x^2 + \rho_v^{\pm 1}}[\sum_{\xi = \pm} \alpha_\xi \sqrt{x^2 + \rho_v^{\xi}}]^2}.
\end{align}
\end{subequations}
The non-trivial coefficients of the RG equations are due to the presence of an unequal number of Dirac excitations that have distinct velocities and coupling constants (i.e. $\rho_v\neq 1$ and $\rho_g\neq 1$).
Note that in the isotropic limit $\rho_v = \rho_g  = 1$ we find that all these functions reduce to $f_{1,2}^\pm = f_3=1$
and we  reproduce results from the literature.\cite{RosensteinPark1991,KhveshchenkoPaaske2001,MosheZinnJustin2003}

\begin{figure}[t]
    \centering
    \includegraphics[scale =1]{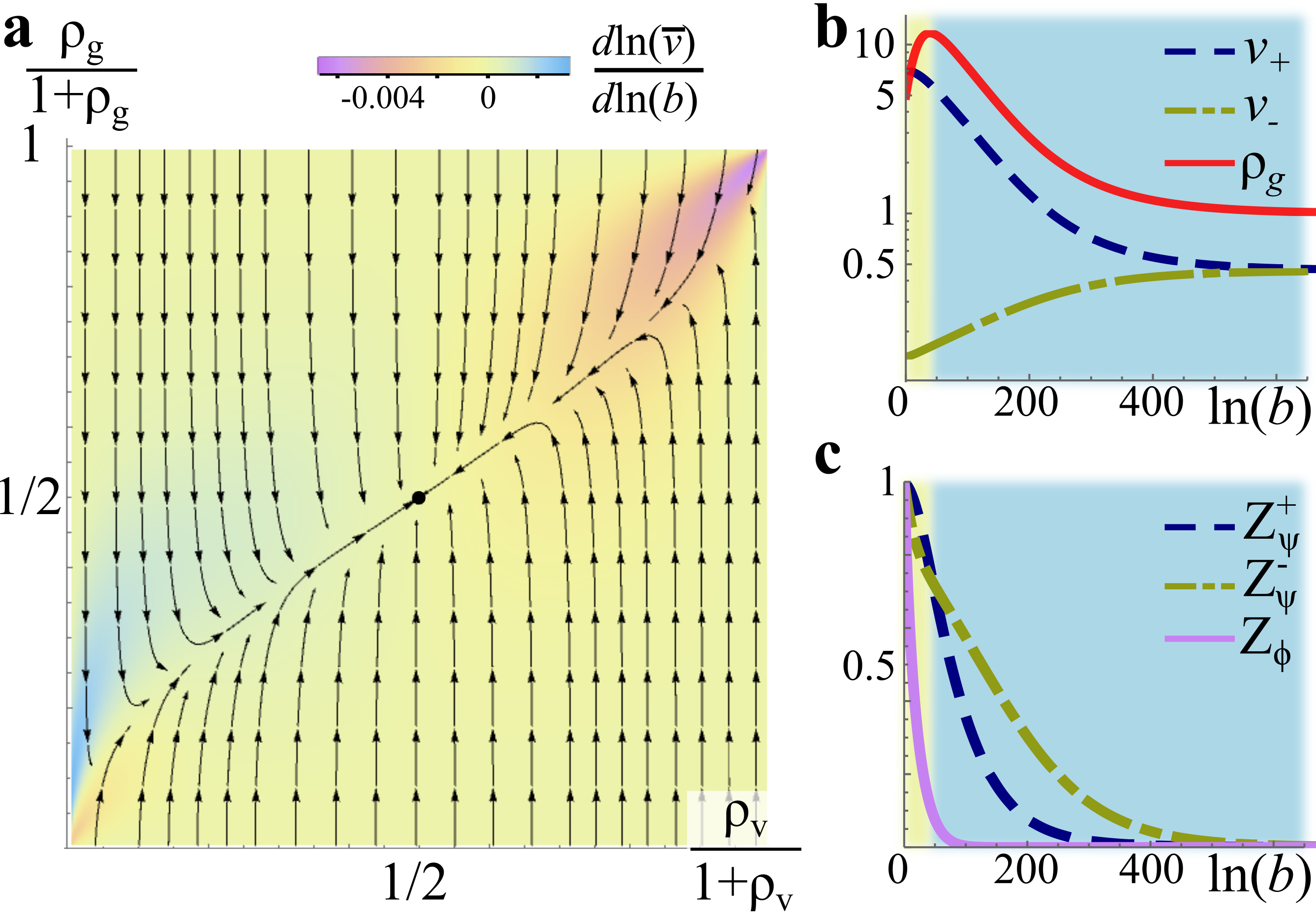}
    \caption{a) RG Flow, Eqs.~\eqref{eq:RGEquations} in the plane of ratios of velocities $\rho_v = v_+/v_-$ and couplings $\rho_g = g_+/g_-$, The color coding represents the beta function of the average velocity $\bar v = \sqrt{v_+ v_-}$. The apparent asymmetry under $\rho_{v,g} \leftrightarrow 1/\rho_{v,g}$ vanishes\cite{SuppMat} in the limit $N_- = N_+$.  b) Integration of RG equations Eq.~\eqref{eq:BetaRhog}-\eqref{eq:Betav} for starting values $\rho_v = 50, \rho_g = 5, \bar v = 1$. c) Field renormalizations, Eq.~\eqref{eq:ZPsi}-\eqref{eq:ZPhi}.
   In panels b), c) we indicate the first (second) stage of RG by yellow (blue) background color. In all panels we set $N_- = 2N_+ = 8$, as in TTG. 
    }
    \label{fig:RGFlow}
\end{figure}

\PRLsection{Analysis of RG flow.} The renormalization group equations of the ratio of velocities and coupling constant are closed, i.e.~the right hand side of Eqs.~\eqref{eq:BetaRhog},\eqref{eq:BetaRhov} only depends on the two running coupling constants $\rho_g, \rho_v$ themselves. 
We solve these equations numerically and present the solution in Fig.~\ref{fig:RGFlow} a). 
We find an attractive fixed point $\rho_v = \rho_g = 1$, where Lorentz symmetry is emergent and two lines of repulsive fixed points where only one of the two fermion species couples to the critical boson, $\rho_{g} \in \lbrace 0, \infty \rbrace$. In particular, the two fixed points at $\rho_{v} = 0 = \rho_g$ and $\rho_{v} = \infty = \rho_g$ are connected to the attractive fixed point by a separatrix. Generally, during the evolution the two coupling constants, $\rho_{v,g}$ approach this separatrix in a first, quick RG stage, and then slowly approach isotropy. 
This shows that, independent on the difference between bare velocities and coupling constants, the fast and slow Dirac fermions cannot be considered separately once correlation effects are taken into account. 

To describe the RG flow for TTG, we choose $N_- = 2 N_+$ and starting values $\rho_v \gg 1$, $\rho_g \sim 1$ on the right in Fig.~\ref{fig:RGFlow} a). 
We Taylor expand the RG equations about this limit using numerically evaluated series coefficients, and obtain
\begin{align}
\frac{d \ln \rho_g}{d \ln b} &\approx \frac{4}{N \pi^2}, \qquad
\frac{d \ln m}{d \ln b} \approx 1 - \frac{8}{N\pi^2} - 12 \frac{\rho_g - 1}{N \pi^2},\nonumber\\
\frac{d\ln Z_{\psi}^-}{d\ln b} &\approx -\frac{1}{N\pi^2},\quad
\frac{d\ln Z_{\phi}}{d\ln b} \approx -\frac{4}{N\pi^2},
\end{align}
while 
\begin{align}
\frac{d \ln \rho_v}{d \ln b} = 
\frac{d \ln \bar v}{d \ln b} = 
\frac{d\ln Z_{\psi}^+}{d\ln b} = \mathcal O(1/\rho_v).
\end{align}
Thus $\rho_v$ and $\bar v$ are only weakly affected in comparison to $\rho_g$, because the RG leads to a massive enhancement of $\rho_g$. 
However, we can still observe that $v_-$ gets enhanced and $v_+$ reduced (Fig.~\ref{fig:RGFlow} b). 
We also see that all field renormalizations start to decrease so that bosonic and fermionic excitations become less coherent, Fig.~\ref{fig:RGFlow} c). 
This first stage of RG, illustrated by a yellow background in Fig.~\ref{fig:RGFlow} b) and c), lasts until $\rho_g(b) \sim \rho_v(0)$ corresponding to length scales below $L_{\rm crossover} \sim [\rho_v(0)/\rho_g(0)]^{N \pi^2/4}/\Lambda$,   
which can be very large. 

In the second stage of RG, depicted by a blue background in Fig.~\ref{fig:RGFlow} b) and c), both $\rho_{v,g}$ very slowly approach unity and there is a moderate renormalization of $\bar v$. 
Of course, if starting values of $\rho_g(0) \simeq \rho_v(0)$, the system may only explore the second regime in which the coupling constants slowly approach $\rho_v = \rho_g = 1$. 
In the vicinity of the attractive fixed point, the flow along the separatrix, which is asymptotically defined by $\rho_g = (3\rho_v-1)/2$, is characterized by a scaling dimension $-{8}/({15 \pi ^2 N}) \stackrel{N = 6}{\approx} -0.009$.  
corresponding to the least irrelevant operator. 
At the attractive fixed point, the average velocity $\bar v$ approaches a finite (non-universal) value, which is determined by the relative difference between fast and slow flavors $(N_+-N_-)/(N_++N_-)$\cite{SuppMat}. Thus for TTG, where $N_->N_+$, $\bar v$ decreases.

\PRLsection{Observables and consequences for experiments.} Observables which allow to experimentally access the velocity renormalization include quantum oscillations, as exploited to uncover the velocity renormalization in suspended graphene\cite{EliasGeim2011}. Another probe is the tunneling density of states that within our RG calculation is given by
\begin{equation}
    \nu(E) =  \sum_\pm N_\pm\frac{Z_{\psi}^\pm(b_E) }{2\pi v_\pm ^2(b_E)}|E|.
\end{equation}
It not only measures the velocity renormalization, but also the suppression of the quasi-particle weight at the corresponding energy scale $b_E = \bar v\Lambda/ E$, see Fig.~\ref{fig:RGFlow} c.

Based on our estimate for the extent of the first stage of the RG flow, we expect that experiments mainly explore this first regime. 
For example, for transport experiments, the maximum RG time can be estimated as $\ln(b_{\rm final})- \ln(b_{\rm initial}) \sim \ln(500 K/10 mK)\sim 10$, where $50 meV \sim 500 K$ is the mini bandwidth of the fast bands and $10 mK$ the base temperature of dilution refrigerators. 

At the same time, 
the bare coupling constants are  a priori  unknown and variations in twist angle affect the bare $\rho_v$ in TTG. In addition, other Dirac materials also have different starting values. In these situations
other segments of the RG flow can be probed. This is the rationale behind the above complete theoretical analysis of the RG flow. 
Assuming starting values as in Fig.~\ref{fig:RGFlow} b), we highlight that in the initial, in TTG experimentally accessible regime $\ln(b) \in (0,10)$, $\rho_g$ doubles and $\rho_v$ increases by 20 \% - both are certainly measurable effects. 

Another implication of our analysis regards the quantum-critical regime as the system is tuned across the quantum phase transition, where experiments can probe quantum critical fluctuations akin to what has been observed in strongly correlated materials. To this end, due to the parametrically slow RG flow, we define 'local' scaling exponents for the correlation length and anomalous dimensions via $1/\tilde\nu=d \ln m/d\ln b$, $\tilde\eta_{\psi \pm}=-d\ln Z_{\psi}^\pm/2d\ln b$ and $\tilde\eta_\phi=1-d\ln Z_\phi/d\ln b$ using Eqs.~\eqref{eq:Betam}-\eqref{eq:ZPhi}. These are not the standard universal critical exponents but scale-dependent quantities, unless the RG flow reaches one of the fixed points. 
However, they establish an analytical relationship between the velocity renormalization and experimentally observable exponents. 
At the attractive Lorentz-symmetric fixed point, they reduce to the universal scaling exponents of the Ising GN model\cite{RosensteinPark1991, KhveshchenkoPaaske2001}: 
$\nu = 1 + 8/(3\pi^2 N)$, $\eta_{\psi+}=\eta_{\psi-}=1/(3\pi^2 N)$, $\eta_\phi=1-16/(3\pi^2 N)$.

\PRLsection{Effect of displacement field and Dirac node offset.} A major experimental tuning knob in twisted trilayer graphene is the displacement field $D_0$. The displacement field generates a hybridization of order $D_0$ of fast and slow modes located at the mini-$K$ point. This hybridization vertically splits the two Dirac points to energies $\pm D_0$ and generates a Fermi surface of size $p_F = D_0/\sqrt{v_+v_-}$ at charge neutrality.
The renormalization group flow calculated above describes the system at energies larger than $D_0$. The evaluation of running coupling constants at the scale $D_0$ serves as bare values for the physics at lower energies, which is not of Gross-Neveu type (instead, e.g., logarithmic Fermi surface instabilities may occur in Cooper and density wave channels).

We also comment on a possible offset in energy between the Dirac node of fast and slow fermions. Such an offset would also provide an infrared cut-off for the integral of RG equations, beyond which the impact of the Fermi surface in one of the two carrier types becomes relevant. \textit{Ab initio}\cite{ParkJarillo2021} calculations for TTG predict an offset of the order of meV which is thus small as compared to the mini-band width.

\PRLsection{Conclusions:} 
Motivated by the recent discoveries in TTG,
we have presented the leading renormalization group equations for an Ising Gross-Neveu theory in $D = 2+1$ dimensions where the order parameter field couples to $N_+$ fast and $N_-$ slow fermion fields. We found a two-stage RG flow, with an ultimate Lorentz-symmetric fixed point at which all fermion velocities are equal, but the mean velocity is reduced, thereby enhancing correlation physics. 
We predict that this velocity renormalization leads to an increase of the slow and decrease of the fast Dirac velocity in TTG. 
We remark that a U(1) generalization of our Gross-Neveu theory would correspond to intervalley coherent order in TTG~\cite{ChristosScheurer2021}, which is expected to display qualitatively similar behavior to the velocity renormalization and critical exponents we have found based on studies of the XY-Gross-Neveu model\cite{ROSENSTEIN1993381,Roy2016,PhysRevD.96.096010,PhysRevB.94.205106,Li2017,PhysRevB.96.115132,Iliesiu2018}. The generalization of our theory to experimentally relevant\cite{park2021magicangle,ZhangNadjPerge2021} twisted pentalayar graphene and other van-der-Waals materials with more than two Dirac velocities is left for future work.
Beyond our application to TTG at charge neutrality, where $N_+ = 4$ and $N_- = 8$, we highlight that our theory may also apply to TTG in the vicinity of correlated insulating states at integer filling per moir\'e unit cell. Following the paradigm of a cascade\cite{Zondiner2020,WongYazdani2020} of resetting Dirac node energies, the effective theory may again be described by Eq.~\eqref{eq:L0}, yet with a reduced number of slow modes $N_-<8$.

\PRLsection{Acknowledgments:} It is a pleasure to thank Lukas Janssen, Pablo Jarillo-Herrero, Eslam Khalaf, Walter Metzner, Pavel Ostrovsky for useful discussions. 
JHP is partially supported by the Air Force Office of Scientific Research under Grant No.~FA9550-20-1-0136 and the Alfred P. Sloan Foundation through a Sloan Research Fellowship.
JHP and EJK acknowledge hospitality by the Aspen Center for Physics, where part of this work was completed and which is
supported by National Science Foundation grant PHY1607611. Work at BNL is supported by the U.S. Department of Energy (DOE), Office of Basic Energy Sciences, under Contract No. DE- SC0012704.

\bibliography{TTG_RG}

\clearpage
\begin{widetext}

\begin{center}
Supplementary materials on \\
\textbf{``Interaction induced velocity renormalization in magic angle twisted trilayer graphene''}\\
Laura Classen$^{1,2}$, J.H. Pixley$^{3}$, Elio J. K\"onig$^{2}$\\ 
$^{1}$ \textit{Condensed Matter Physics and Materials Science Division, Brookhaven National Laboratory, Upton, New York 11973, USA}\\
$^{2}$ \textit{Max Planck Institute for Solid State Research, D-70569 Stuttgart, Germany}\\
$^{3}$\textit{Department of Physics and Astronomy, Center for Materials Theory, Rutgers University, Piscataway, NJ 08854}
\end{center}
\end{widetext}

\setcounter{equation}{0}
\setcounter{figure}{0}
\setcounter{section}{0}
\setcounter{table}{0}
\setcounter{page}{1}
\makeatletter
\renewcommand{\theequation}{S\arabic{equation}}
\renewcommand{\thesection}{S\arabic{section}}
\renewcommand{\thefigure}{S\arabic{figure}}

These supplementary materials contain a motivation of our model,~Sec.~\ref{SM:sec:EffModel}, in connection to twisted trilayer graphene, and the explicit calculations for mean field solution, Sec.~\ref{app:MF}, and RG equations, Sec.~\ref{app:LoopCorrections}, stated in the main text, as well as an analysis of the RG flow, Sec.~\ref{SM:sec:RGFlow}.

\section{Derivation of effective model}
\label{SM:sec:EffModel}
We consider symmetrically twisted trilayer graphene in zero displacement field near charge neutrality, which can be described by (uncoupled) fast and slow Dirac fermions on the single-particle level. We assume that interactions lead to an instability in a particle-hole channel with Ising symmetry. 
In that case, 
the most general attraction between different fermion species in the interaction channel under consideration is
\begin{align}
    \mathcal L_{\rm int} &= - \sum_{\xi, \xi' = \pm 1}\frac{g_{\xi \xi'}^2}{2N} (\bar \psi_\xi \sigma_z \psi_\xi)(\bar \psi_{\xi'} \sigma_z \psi_{\xi'}) \notag\\
    &\doteq \frac{N}{2} \vec \phi^T \left ( \begin{array}{cc}
        g_{++}^2 & g_{+-}^2 \\
        g_{+-}^2 & g_{--}^2
    \end{array}\right)^{-1} \vec \phi \notag \\
    & + \left (\bar \psi_+ \sigma_z \psi_+,\bar \psi_- \sigma_z \psi_- \right) \vec \phi. \label{eq:HintAppendix}
\end{align}
Here, we have introduced the Hubbard-Stratonovich field $\vec \phi = (\phi_+, \phi_-)^T$, which is composed of the order parameter field in fast and slow sectors and we suppress the flavor index $\alpha$.
Importantly, the order parameter fields in fast and slow sectors have the same symmetry properties, such that the microscopic interaction generally couples them, i.e. $g_{+-} \neq 0$.

In the large N limit, the instability occurs when the inverse RPA propagator
\begin{equation}
    D^{-1}(\omega = 0, \vec q = 0) = \left ( \begin{array}{cc}
        g_{++}^2 & g_{+-}^2 \\
        g_{+-}^2 & g_{--}^2
    \end{array}\right)^{-1} + \frac{1}{N} \left ( \begin{array}{cc}
        \Pi_+ & 0 \\
        0 & \Pi_-
    \end{array} \right)
\end{equation}
acquires a zero mode (here $\Pi_{\pm}$ is the fermionic susceptibility at zero external frequency and momentum). One may expand $\vec \phi = \phi \hat e_0 + \phi_\perp \hat e_\perp$, where $D^{-1}(\omega = 0, \vec q = 0)\hat e_0 = 0$ at criticality, while $D^{-1}(\omega = 0, \vec q = 0)\hat e_\perp = C \hat e_\perp$, $C>0$. Thus $\hat e_0 \hat e_0^T$ is the projector on the interaction channel which first acquires a zero mode, as determined by the condition $\det[D^{-1}(\omega = 0, \vec q = 0)] = 0$ or
\begin{equation}
    1 + (g_{++}^2 \Pi_{+}+g_{--}^2 \Pi_{-})/N = (g_{+-}^4-g_{++}^2g_{--}^2)\Pi_{+}\Pi_{-}/N^2.
\end{equation}
Clearly, this condition reduces the number of independent coupling constants $g_{++},g_{--},g_{+-}$ from three to two. 

In the main text, we tacitly project the interaction onto the critical channel $\hat e_0$ and drop the massive modes $\hat e_\perp$ altogether (the impact of virtual fluctuations can be assumed to be weak without changing qualitative features of the theory). Technically, the projection amounts to keeping only fluctuations $\vec \phi =\phi \hat e_0$ in Eq.~\eqref{eq:HintAppendix}, i.e. 
\begin{align}
      \mathcal L_{\rm int} &= \frac{N \phi^2}{2} \hat e_0^T \left ( \begin{array}{cc}
        g_{++}^2 & g_{+-}^2 \\
        g_{+-}^2 & g_{--}^2
    \end{array}\right)^{-1} \hat e_0 \notag \\
    & + \phi \left (\bar \psi_+ \sigma_z \psi_+,\bar \psi_- \sigma_z \psi_- \right) \hat e_0.
\end{align}

We define $\bar g^{-2} = \hat e_0^T \left ( \begin{array}{cc}
        g_{++}^2 & g_{+-}^2 \\
        g_{+-}^2 & g_{--}^2
    \end{array}\right)^{-1} \hat e_0$ and absorb $\phi \rightarrow \bar g \phi$. Then, by comparison with Eq.~\eqref{eq:L0} of the main text, $g_+ = \bar g (1,0)^T \hat e_0$, $g_- = \bar g (0,1)^T \hat e_0$.

We conclude this section with an estimate of $g_+$, $g_-$ in terms of $g_{++},g_{+-},g_{--}$ in the important limit $\vert \Pi_{++} \vert \ll \vert \Pi_{--} \vert \sim g_{++}\sim g_{+-} \sim g_{--}$. Criticality is driven by slow modes, leading to $\hat e_0 = (g_{-+}^2, g_{--}^2)/\sqrt{g_{-+}^4 + g_{--}^4}$, $\hat e_\perp = (g_{--}^2,-g_{-+}^2)/\sqrt{g_{-+}^4 + g_{--}^4}$, with the approximate mean field transition at $1 + g_{--}\Pi_-/N = 0$. By consequence, we can estimate the parameters of Eq.~\eqref{eq:L0} of the main text as $g_+ = g_{-+}^2/g_{--}, g_- = g_{--}$.

\section{Model and mean field solution}
\label{app:MF}

As outlined in the main text and in the previous section of this supplement, we consider the model
\begin{equation}
\mathcal L = \sum_{\pm} \sum_{\alpha = 1}^{N_\pm} \bar \psi_{\alpha, \pm} [\partial_\tau + v_\pm \vec p \cdot \vec \sigma + g_\pm \phi \sigma_z] \psi_{\alpha, \pm} + \frac{N}{2} \phi^2.
\end{equation}
We assume $N_+ + N_- = 2N \gg 1$. It is convenient to parametrize
\begin{equation}
g_\pm = \bar g \rho_g^{\pm 1/2}, \quad v_\pm = \bar v \rho_v^{\pm 1/2},
\end{equation}
where $\bar g = \sqrt{g_+ g_-}$ and $\rho_g = g_+/g_-$ and analogously for velocity. We absorb $\bar g$ into $\phi$, then

\begin{equation}
\mathcal L = \sum_{\pm} \sum_{\alpha = 1}^{N_\pm} \bar \psi_{\alpha, \pm} [\partial_\tau + v_\pm \vec p \cdot \vec \sigma + \rho_g^{\pm 1/2} \phi \sigma_z] \psi_{\alpha, \pm} + \frac{N}{2 \bar g^2} \phi^2. \label{eq:L0RG}
\end{equation} 

In the case, $N_+ = N_- = N$ and $v_+ = v_- = \rho_g = 1$, this model is the same as the Gross-Neveu(-Yukawa) model, see Ref.~\onlinecite{RosensteinPark1991} of the main text. 
Note that the energy dimension of $\phi$ and $\psi$ is one, and of $\bar g^2$ is -1. 

\subsection{Fermionic Green's function}

When the $\phi$ field condenses, the inverse fermionic Green's functions are 
\begin{equation}\label{eq:FermionicGreensFunction}
G_\pm^{M_\pm}(\v p)^{-1} = (i \epsilon - v_\pm \vec p \cdot \vec \sigma  - M_\pm\sigma_z),
\end{equation}
where $M_\pm = \rho_g^{\pm 1/2} \langle \phi \rangle \equiv  \rho_g^{\pm 1/2} \bar M$. We use the notation of a bold symbol $\mathbf p = (\epsilon, \vec p)$ for energy-momentum vectors. Inverting the matrix yields
\begin{equation}
   G_\pm^{M_\pm}(\v p)=-\frac{i\epsilon+v_\pm \vec p\cdot \vec \sigma+M_\pm \sigma_z}{\epsilon^2+v_\pm^2\vec p^2 +M_\pm^2} .
\end{equation}
We use non-zero $M_\pm$ to simplify some calculations, but generally we are interested in the symmetric phase and consider $M_\pm\rightarrow 0$. 

\subsection{Polarization operator and effective boson propagator}

The polarization operator (or susceptibility), Fig.~\ref{fig:Diagrams} a) of the main text, of each fermionic species is
\begin{eqnarray}
\Pi_\pm(\v q) &=&  N_\pm \rho_g^{\pm 1} \int_{\v p} \tr^\sigma[G_\pm^{M_\pm}(\v p + \v q/2)G_\pm^{M_\pm}(\v p - \v q/2)] \notag\\
&=& \frac{N_\pm}{\pi}\frac{\rho_g^{\pm 1}}{v_\pm^2} \Big [ - ({v_{\pm} \Lambda/2} -  M_\pm/2) \notag \\
&&+ \frac{4 M_\pm^2 + q_\pm^2}{4 q_\pm} \arctan\left (\frac{q_\pm}{2M_\pm} \right) \Big].
\end{eqnarray}
We introduced the notation $q_\pm = \sqrt{\omega^2 + v_{\pm}^2 \vec q^2}$, the ultraviolet cut-off $\Lambda$ ($\vec p^2 < \Lambda^2$) and $\v q = (\omega, \vec q)$, $\v p = (\epsilon, \vec p)$.
Thus, to leading order in $N \gg 1$ we get the effective bosonic theory 
\begin{equation}
\label{eq:Leffboson}
\mathcal L_{\rm eff}[\phi] = \frac{N}{2} \int_{\v q} \phi(-\v q) \underbrace{[\frac{1}{ \bar g^2} + \frac{1}{N}\sum_\pm \Pi_\pm(\v q)]}_{\equiv D^{-1}(\v q)/N} \phi(\v q).
\end{equation}
We see that, in the symmetric phase where $\bar M = 0$, the effective mass of bosons is given by
\begin{align} \label{eq:BosonicMass}
m 
& = \frac{1}{\bar g^2} - \frac{\Lambda}{2\pi \bar v }\sum_\pm \alpha_\pm \rho_v^{\pm 1/2},
\end{align}
where $
\alpha_\pm = \frac{N_\pm}{N} \frac{\rho_g^{\pm 1} }{\rho_v^{\pm 1}}$. We stated this in Eq.~\eqref{eq:BosonicMassMainText} of the main text.

$m =0$ defines the mean field transition. We remark that in Ref.~\onlinecite{RosensteinPark1991} of the main text, the UV cutoff is defined as ${\frac{\bar v \pi \Lambda}{2}}={\Lambda_{\rm Rosenstein}}$. From Eq.~\eqref{eq:Leffboson}, we obtain the effective bosonic propagator 
\begin{equation}
D(\v q) = \frac{1}{N} \frac{1}{m + \sum_\pm \alpha_\pm q_\pm/(8 \bar v^2)}.
\end{equation}
The dimension of the propagator is 1/energy. 
At the isotropic point we can set $v_+ = v_-$ and $\alpha_\pm = 1$.

\begin{figure}[b]
    \centering
    \includegraphics[width = .45\textwidth]{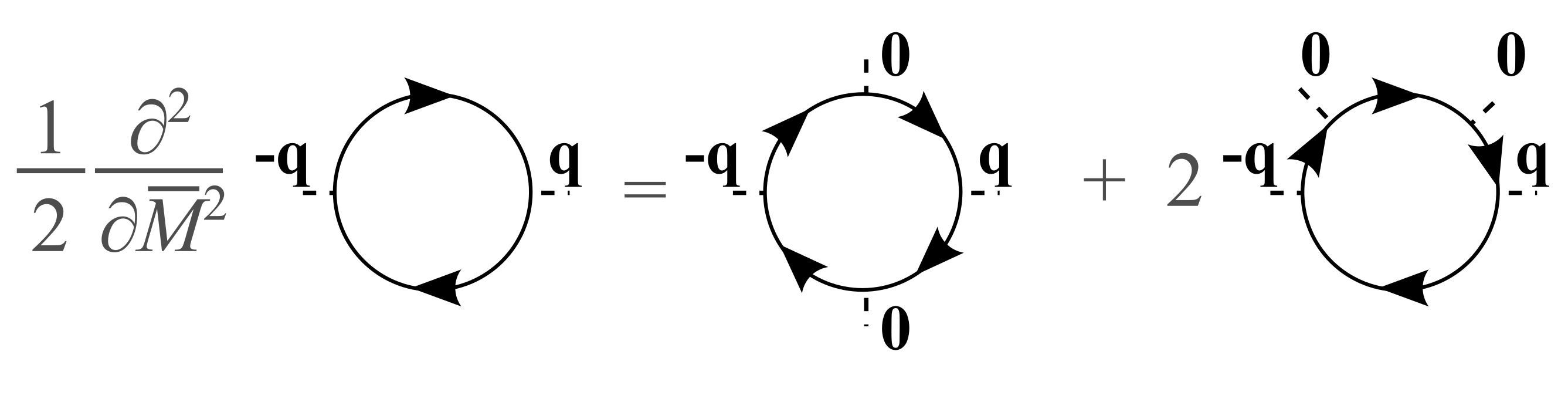}
    \caption{Diagrammatic representation of Eq.~\eqref{eq:PiExp}. The differentiation with respect to $\bar M$ introduces two external legs at zero momentum. In a subsequent step, to calculate the diagrams of Fig.~\ref{fig:Diagrams} d), e) of the main text, the legs at finite $\v q$ are contracted. This leads to the loop correction of the zero momentum susceptibility.}
    \label{fig:Diagrams2}
\end{figure}

For later reference, we also remark the following relations here. In 
the calculation of the fermionic self-energy we need 
\begin{equation}
D(\v q + \v p) \stackrel{\v p \ll \v q}{\simeq} D(\v q)\left (1 - D(\v q) \frac{N}{4 \bar v^2}\sum_\pm \frac{\alpha_\pm}{2} \frac{\langle \v p, \v q\rangle_\pm}{q_\pm} \right), \label{eq:DqExp}
\end{equation}
with the scalar product
$\langle  \v p, \v q \rangle_\pm = \epsilon \omega + v_\pm^2 \vec p \cdot \vec q$.
For the calculation of the two-loop correction to the bosonic mass, Fig.~\ref{fig:Diagrams} d),e) of the main text, we calculate the second derivative of the polarization operator with respect to the mass $\bar M = \langle \phi \rangle$, see Fig.~\ref{fig:Diagrams2}. We find

\begin{eqnarray}
\frac{1}{2} \frac{\partial^2}{\partial \bar M^2} \Pi_\pm(\v q) &=&  \frac{\rho_g^{\pm 1}}{2} \frac{\partial^2}{\partial M_\pm^2} \Pi_\pm(\v q) \notag \\
&=& \frac{N_\pm \rho_g^{\pm 2}}{\pi v_\pm^2 q_\pm} \left (\arctan\left (\frac{q_\pm}{2M_\pm}\right)  - \frac{2M_\pm q_\pm}{q_\pm^2 + 4 M_\pm^2}\right) \notag \\
&\simeq & N \frac{\rho_g^{\pm 1}}{\bar v^2} \frac{\alpha_\pm}{2 q_\pm} . \label{eq:PiExp}
\end{eqnarray}

\section{Loop Corrections and renormalization group}
\label{app:LoopCorrections}

We here present calculations of the loop corrections to the fermionic, and bosonic self energy, as well as details on the RG procedure.

\subsection{Fermionic self energy and vertex correction}

The fermionic self energy and the vertex correction, Fig.~\ref{fig:Diagrams} b) and c) of the main text, are most efficiently calculated by studying the self energy in the symmetry-broken phase
\begin{align}
\Sigma_{\pm}(\v p) &= \int_{\v q} D(\v q- \v p)\sigma_z G_\pm(\v q) \sigma_z \rho_g^{\pm 1} \notag \\
&= - \int_{\v q} \frac{D(\v q)}{q_\pm^2 + M_\pm^2} M_\pm \sigma_z \rho_g^{\pm 1}  \notag \\
&- \int_{\v q} \frac{D(\v q)^2 (i \omega - v_\pm \vec q \cdot \vec\sigma)}{q_\pm^2 + M_\pm^2} \rho_g^{\pm 1} \sum_{\xi = \pm 1}
{N \alpha_\xi/8 \bar v^2} 
\frac{\langle \v p, \v q \rangle_\xi}{q_\xi} \notag\\
&= - I_1^\pm M_\pm \sigma_z \rho_g^{\pm 1} - I_2^\pm i \epsilon \rho_g^{\pm 1} + I_3^\pm \rho_g^{\pm 1} v_\pm \vec p \cdot \vec \sigma, 
\label{eq:SEcorrect}
\end{align}
where we repeat from the main text that 
$\alpha_\pm = N_\pm \rho_g^{\pm 1}/(N \rho_v^{\pm 1})$.
The leading logarithmic contributions are encoded in the following dimensionless integrals
\begin{align}
I_1^\pm &= \int^\Lambda \frac{d^2 q}{(2\pi)^2} \int \frac{d\omega}{2\pi} \frac{D(\v q)}{q_\pm^2},\\
I_2^\pm &= \frac{N}{8 \bar v^2}\int^\Lambda \frac{d^2 q}{(2\pi)^2} \int \frac{d\omega}{2\pi} \frac{D(\v q)^2}{q_\pm^2} \sum_\xi \alpha_\xi \frac{\omega^2}{q_{\xi}} ,\\
I_3^\pm &= \frac{N}{8 \bar v^2}\int^\Lambda \frac{d^2 q}{(2\pi)^2} \int \frac{d\omega}{2\pi} \frac{D(\v q)^2}{q_\pm^2} \sum_\xi \alpha_\xi \frac{1}{2} \frac{v_\xi^2 \vec q^2}{q_{\xi}}.
\end{align}
To evaluate these integrals, we change the variables $\omega = \bar v x \vert \vec q \vert$, so that we obtain
\begin{align}
I_1^{\pm} &= \frac{2}{N \pi^2} \ln \left ( \frac{\Lambda}{m \bar v}\right) g_1^{\pm} (v_+, v_-, g_+, g_-, N_+, N_-), \\
I_2^{\pm} &= \frac{2}{3 N \pi^2} \ln \left ( \frac{\Lambda}{m \bar v}\right) g_2^{\pm} (v_+, v_-, g_+, g_-, N_+, N_-), \\
I_3^{\pm} &= \frac{2}{3 N \pi^2} \ln \left ( \frac{\Lambda}{m \bar v}\right) g_3^{\pm} (v_+, v_-, g_+, g_-, N_+, N_-),
\end{align}
where all of the following terms are unity in the isotropic limit
\begin{align}
g_1^{\pm} &= \int_{-\infty}^\infty dx \frac{1}{({x^2 + \rho_v^{\pm 1}})[\sum_\xi \alpha_\xi \sqrt{x^2 + \rho_v^{\xi}}]},\\
g_2^{\pm} &= 3 \int_{-\infty}^\infty dx \frac{\sum_{\xi = \pm} \alpha_\xi \frac{x^2}{\sqrt{x^2 + \rho_v^{\xi}}}}{({x^2 + \rho_v^{\pm 1}})[\sum_{\xi = \pm} \alpha_\xi \sqrt{x^2 + \rho_v^{\xi}}]^2},\\
g_3^{\pm} &= 3 \int_{-\infty}^\infty dx \frac{\sum_{\xi = \pm} \alpha_\xi \frac{\rho_v^\xi/2}{\sqrt{x^2 + \rho_v^{\xi}}}}{({x^2 + \rho_v^{\pm 1}})[\sum_{\xi = \pm} \alpha_\xi \sqrt{x^2 + \rho_v^{\xi}}]^2}.
\end{align}

Note that none of $g_{1,2,3}^\pm$ depends explicitly on $\bar g$ or $\bar v$ and that $g_2^\pm = 3 g_1^\pm - 2 g_3^\pm$. In the isotropic case $\rho_v=\rho_g=1$, we get $g_1^\pm=g_2^\pm=g_3^\pm=1$. We also remark that we dropped terms with logarithmically slow dependence on $\rho_v, \rho_g$.

\subsection{Two-loop correction to bosonic self-energy}

We calculate the leading correction to the polarization operator at zero external momentum, Fig.~\ref{fig:Diagrams} d) and e) of the main text, by  
exploiting Eq.~\eqref{eq:PiExp} and obtain

\begin{align}
    \delta \Pi_\pm(0) &= \int_{\v q} D(\v q)  N \frac{\rho_g^{\pm 1}}{\bar v^2} \frac{\alpha_\pm}{2 q_\pm} \notag \\
    &= \frac{1}{8 \pi^2 \bar v^2} \int dx  \frac{\rho_g^{\pm 1} \alpha_\pm}{\sqrt{x^2 + \rho_v^{\pm 1}}} \notag \\
    &\times \int^\Lambda dq \frac{ q}{m + \frac{q}{8 \bar v} \sum_\xi\alpha_\xi \sqrt{x^2 + \rho_v^\xi}} \notag \\
    &= \frac{\Lambda}{2\pi \bar v} \rho_g^{\pm 1} \alpha_\pm g_4^\pm - m \frac{4 \rho_g^{\pm 1} \alpha_\pm}{\pi^2} g_5^\pm\ln\left (\frac{\Lambda}{m \bar v}\right).
\end{align}

 Again, we dropped slow, logarithmic dependence on $\rho_v, \rho_g$ and introduced two more functions which are unity in the isotropic limit
\begin{align}
    g_4^\pm &= \int_{-\infty}^\infty dx \frac{2/\pi}{\sqrt{x^2 + \rho_v^{\pm 1}} \sum_\xi \alpha_\xi \sqrt{x^2 + \rho_v^\xi}},\\
    g_5^\pm &= \int_{-\infty}^\infty dx \frac{2}{\sqrt{x^2 + \rho_v^{\pm 1}} (\sum_\xi\alpha_\xi \sqrt{x^2 + \rho_v^\xi})^2}.
\end{align}
This concludes the calculation of divergent leading order diagrams.

\subsection{Counter-terms and derivation of RG equations}

We next present details on the derivation of the RG equations. We introduce renormalized quantities to obtain the renormalized action
\begin{align}
\mathcal L &=  \sum_{\pm,\alpha} \bar \psi_{\alpha, \pm}^R [Z_1^\pm \partial_\tau +Z_1^\pm Z_v^\pm v_\pm^R \vec p \cdot \vec \sigma + Z_2^\pm (\rho_g^R)^{\pm 1/2} \phi^R \sigma_z] \psi_{\alpha, \pm}^R \notag \\
&+ \frac{Z_3 Z_2^+ Z_2^-}{Z_1^+ Z_1^-}\frac{N}{2 (\bar g^R)^2} (\phi^R)^2 , \label{eq:RenormAction}
\end{align}
where the renormalization factors $Z_i$ contain counter-terms such that the effective generating functional in terms of renormalized coupling constants and source fields conjugate to renormalized fields (indicated by a superscript $R$) is finite. The counterterms are introduced to render the loop diagrams calculated above finite in the limit $\Lambda \rightarrow \infty$, but come at the expense of  
endowing physical quantities with a non-trivial scaling with respect to a running scale $\mu$.
We can 
read off the renormalization factors curing the one-loop diagrams from Eq.~\eqref{eq:SEcorrect}
\begin{subequations}
\begin{align}
    Z_{1}^\pm &= 1- \ln(\Lambda/\mu) \frac{2}{3 \pi^2 N} g_2^{\pm}  \rho_g^{\pm 1},\\
    Z_v^\pm &= 1 - \ln(\Lambda/\mu) \frac{2}{3 \pi^2 N} [g_3^{\pm}- g_2^\pm] \rho_g^{\pm 1},\\
    Z_2^\pm & = 1 + \ln(\Lambda/\mu) \frac{2}{\pi ^2 N } g_1^\pm \rho_g^{\pm 1}.
\end{align}
\label{eq:Zs}
\end{subequations}

Next, we determine $Z_3$. To this end we consider the bosonic mass
\begin{widetext}
\begin{align}
    \frac{D^{-1}(0)}{N} &= \frac{Z_3 Z_2^+Z_2^-}{Z_1^+Z_1^- [\bar g^R]^2} + \frac{1}{N} \sum_\pm \left [\frac{Z_2^\pm}{Z_1^\pm}\right]^{2}\Pi_\pm(0) \notag \\
    &= \frac{Z_3}{[\bar g^R]^2} - \sum_\pm\frac{\Lambda}{2\pi \bar v} \alpha_{\pm} (\rho_v^{\pm 1/2} - \frac{1}{N} \rho_{g}^{\pm 1} g_4^\pm) \notag\\
    &- m \ln \left (\frac{\Lambda}{\mu}\right) \frac{8 \tilde f_3}{3\pi^2 N}- \frac{2}{3\pi^2 N} \frac{\Lambda}{2\pi \bar v}\ln \left (\frac{\Lambda}{\mu}\right)  \left (\sum_\pm \pm(6 g_1^\pm - 2 g_2^\pm)\rho_g^{\pm 1} \right)\left (\sum_\pm \pm \alpha_\pm \rho^{\pm 1/2} \right) + \mathcal O (\ln(\mu/m)) . \label{eq:massb}
\end{align} 
\end{widetext}
While it will be of no importance for the final RG equation, we remark that all velocities entering this equation are bare velocities $v^{\rm bare} = Z_v^\pm v^R_\pm$ containing infinities in the form of counter terms of $\mathcal O(1/N)$.
We introduced the function
\begin{equation}
    \tilde f_3 = \sum_\pm\frac{3 \alpha_\pm g_5^\pm -  (3 g_1^{\pm} - g_3^\pm)}{2}\rho_g^{\pm 1}.
\end{equation}

For the calculation of the RG equation, it is more convenient to express Eq.~\eqref{eq:massb} using
\begin{align}
    m_{\rm bare} &= \frac{Z_3}{[\bar g^R]^2} - \sum_\pm\frac{\Lambda}{2\pi \bar v} \alpha_{\pm} (\rho_v^{\pm 1/2} - \frac{1}{N} \rho_{g}^{\pm 1} g_4^\pm) \notag\\
&    - \frac{2}{3\pi^2 N} \frac{\Lambda}{2\pi \bar v}\ln \left (\frac{\Lambda}{\mu}\right)  \left (\sum_\pm \pm(6 g_1^\pm - 2 g_2^\pm)\rho_g^{\pm 1} \right) \notag \\
&\times \left (\sum_\pm \pm \alpha_\pm \rho^{\pm 1/2} \right).
\end{align}
Then, Eq.~\eqref{eq:massb} becomes
\begin{align}
    m_{\rm physical} &= m_{\rm bare} - m_{\rm bare} \frac{8 \tilde f_3}{3 \pi^2 N} \ln \left (\frac{\Lambda}{\mu}\right).
\end{align}

\subsection{Renormalization group equations}
Next, we express the renormalizations of $v_\pm, g_\pm$ by renormalization of relative and mean couplings, i.e.

\begin{align}
\frac{d \ln(\bar v)}{d \ln(\mu)} 
&= -\frac{1}{3 \pi^2 N} [g_3^+ \rho_g^{+1} - g_2^+\rho_g^{+1} + g_3^-\rho_g^{-1} - g_2^-\rho_g^{-1}],  \\
\frac{d \rho_v}{d \ln(\mu)}
&= \frac{2}{3 \pi^2 N} [-g_3^+\rho_g^{+1} + g_3^-\rho_g^{-1} + g_2^+\rho_g^{+1}- g_2^-\rho_g^{-1}] \rho_v,\\
\frac{d \rho_g}{d \ln(\mu)} &= \frac{2}{3\pi^2 N} [3g_1^+\rho_g^{+1} -3g_1^-\rho_g^{-1} + g_2^+\rho_g^{+1} - g_2^-\rho_g^{-1} ] \rho_g, \\
\frac{d m}{d \ln(\mu)} &=  - \frac{8 \tilde f_3}{3 \pi^3 N} m.
\end{align}

In the main text, we use the notation $g_1^\pm \rho_g^{\pm 1} = f_1^\pm$, $g_3^\pm \rho_g^{\pm 1} = f_2^\pm$, $\tilde f_3 = 3 f_3 - \sum_\pm (3f_1^\pm - f_2^\pm)/2 $ in the presentation of these RG equations, where we also use $g_2^\pm \rho_g^{\pm 1}= (3 g_1^\pm - 2 g_3^\pm)\rho_g^{\pm 1} = 3f_1^\pm - 2f_2^\pm$.

\subsection{Wilsonian RG}

The leading order $1/N$ corrections to $\rho_v, \rho_g, \bar v$ as well as the scaling dimensions of the fields all stem from one-loop diagrams. Therefore, the corresponding RG equations can be obtained using a Wilsonian RG protocol including and iterative integration of fast modes with momenta $q\in(\mu = \Lambda/b,\Lambda)$. However, the renormalization of the bosonic mass stems from two-loop diagrams, Fig.~\ref{fig:Diagrams} c) of the main text. Conventional wisdom states that the Wilsonian RG may not be applicable to such higher-order corrections, because of potential miscounting of fast degrees of freedom in the phase space of the momentum shell approach. Here we explicitly rederive the RG equations in the Wilsonian formalism and find full agreement with the field theoretical counter-term approach. At the same time, as we believe that the Wilsonian approach is more transparent to the condensed matter readership, we keep it in this supplement despite the redundancy.

The integration of fast modes in the momentum shell readily lead to the effective action
\begin{align}
S_{\rm eff} &=  \int^< d^2 x\int d\tau \Big [\frac{Z_1^+ Z_1^-}{Z_3^W Z_2^+ Z_2^-}\frac{N}{2 \bar g^2} (\phi^<)^2 \notag \\
&+\sum_{\pm,\alpha} \bar \psi_{\alpha, \pm}^< [(Z_1^\pm)^{-1} \partial_\tau +(Z_1^\pm Z_v^\pm)^{-1} v_\pm \vec p \cdot \vec \sigma \notag \\
&+ (Z_2^\pm)^{-1} (\rho_g)^{\pm 1/2} \phi^< \sigma_z] \psi_{\alpha, \pm}^< \Big ]. \label{eq:WilsonRenormAction}
\end{align}
Here, the notation $\int^<d^2x$ implicitly defines the real space correspendent to momentum integrals $\int^< d^2p  \dots = \int d^2 p \Theta (\Lambda/b - \vert \vec p\vert )\dots$. Similarly, fields with superscript $^<$ are slow fields. The renormalization of coupling constants $Z_1^\pm$, $Z_v^\pm$ and $Z_2^\pm$ directly follow from diagrams of the form Fig.~\ref{fig:Diagrams} b),c) of the main text and are given by Eqs.~\eqref{eq:Zs}, keeping in mind that $\mu = \Lambda/b$. The bosonic term has a prefactor
\begin{align}
    \frac{Z_1^+ Z_1^-}{Z_3^W Z_2^+ Z_2^- \bar g^2} &= \frac{1}{\bar g^2} - \frac{\Lambda/b}{2\pi \bar v} \sum_\pm \alpha_\pm (\rho_v^{\pm 1/2} - g_4^\pm\rho_g^{\pm 1}/N)(b - 1) \notag \\
    &- \frac{4 m}{\pi^2 N} \ln(b) \sum_\pm\alpha_\pm \rho_g^{\pm 1} g_5^\pm.
\end{align}

To obtain this quantum correction, we consider diagrams Fig.~\ref{fig:Diagrams} d) and e) of the main text at a fast bosonic energy momentum with a RPA resummation keeping only bubbles of fast fermions. However, since the momentum is fast, inclusion of fast and slow fermions yields the same result, so that effectively we can use the polarization operator introduced above. 

Next, we rescale fields and space time in a manner to reproduce the appearance of the original action
\begin{align}
    \v x &= b \v x_R, \notag\\
    \psi_\pm^R(\v x_R) &= b (Z_1^\pm)^{-1/2} \psi_\pm^<(\v x) \\
        \phi^R(\v x_R) &= b \sqrt{\frac{Z_1^+Z_1^-}{Z_2^+ Z_2^-}} \phi^<(\v x),
\end{align}

to get
\begin{align}
    S &= \int d^2x_R d \tau_R \Big [\frac{N}{2 \bar g^2_R} [\phi^R(\v x_R)]^2 \notag \\
    &+\sum_{\pm} \sum_{\alpha = 1}^{N_\pm} \bar \psi^R_{\alpha, \pm}(\v x_R) \lbrace\partial_{\tau_R} - i v_\pm^R \vec \nabla_{x_R} \cdot \vec \sigma \notag \\
    &+ [\rho_g^R]^{\pm 1/2} \phi^R(\v x_R) \sigma_z \rbrace \psi^R_{\alpha, \pm}(\v x_R) \Big ].
\end{align}
Now the momentum integral associated to $\int d^2 x_R$ runs up to the original cut-off $\Lambda.$ 
The renormalized coupling constants are
\begin{align}
    v_\pm^R &= \left [1 + \ln(b) \frac{2[g_3^\pm - g_2^\pm]\rho_g^{\pm 1}}{3\pi^2 N} \right] v_\pm, \\
    \rho_g^R &= \left [1 - \ln(b)\sum_\pm (\pm 1) \frac{2[g_2^\pm +3 g_1^\pm ]\rho_g^{\pm 1}}{3\pi^2 N} \right] \rho_g, \\
    m^R &= bm \left (1 - \ln(b)\frac{8 \tilde f_3}{3\pi^2 N} \right ),
\end{align}
where the relationship between coupling constant and bosonic mass is
\begin{equation}
    m^R = \frac{1}{[\bar g^R]^2} - \frac{\Lambda}{2\pi \bar v} \sum_\pm \alpha_\pm (\rho_v^{\pm 1/2} - \frac{g_4^\pm \rho_g^{\pm 1}}{N}).
\end{equation}

The renormalization group equations presented in the main text follow directly from these equations by means of differentiation with respect to $\ln(b)$. To get the renormalization of fields, we define $Z_\psi^\pm = Z_1^\pm$ and $Z_\phi = \prod_\pm Z_2^\pm/Z_1^\pm$. Note that the renormalization group equations reproduce the results obtained in the previous section on the basis of the field theoretical counter-term method.

\section{Analysis of RG flow}
\label{SM:sec:RGFlow}

In this section we present additional details on the analysis of the RG flow.

\begin{figure}[b]
    \centering
    \includegraphics[width = .45\textwidth]{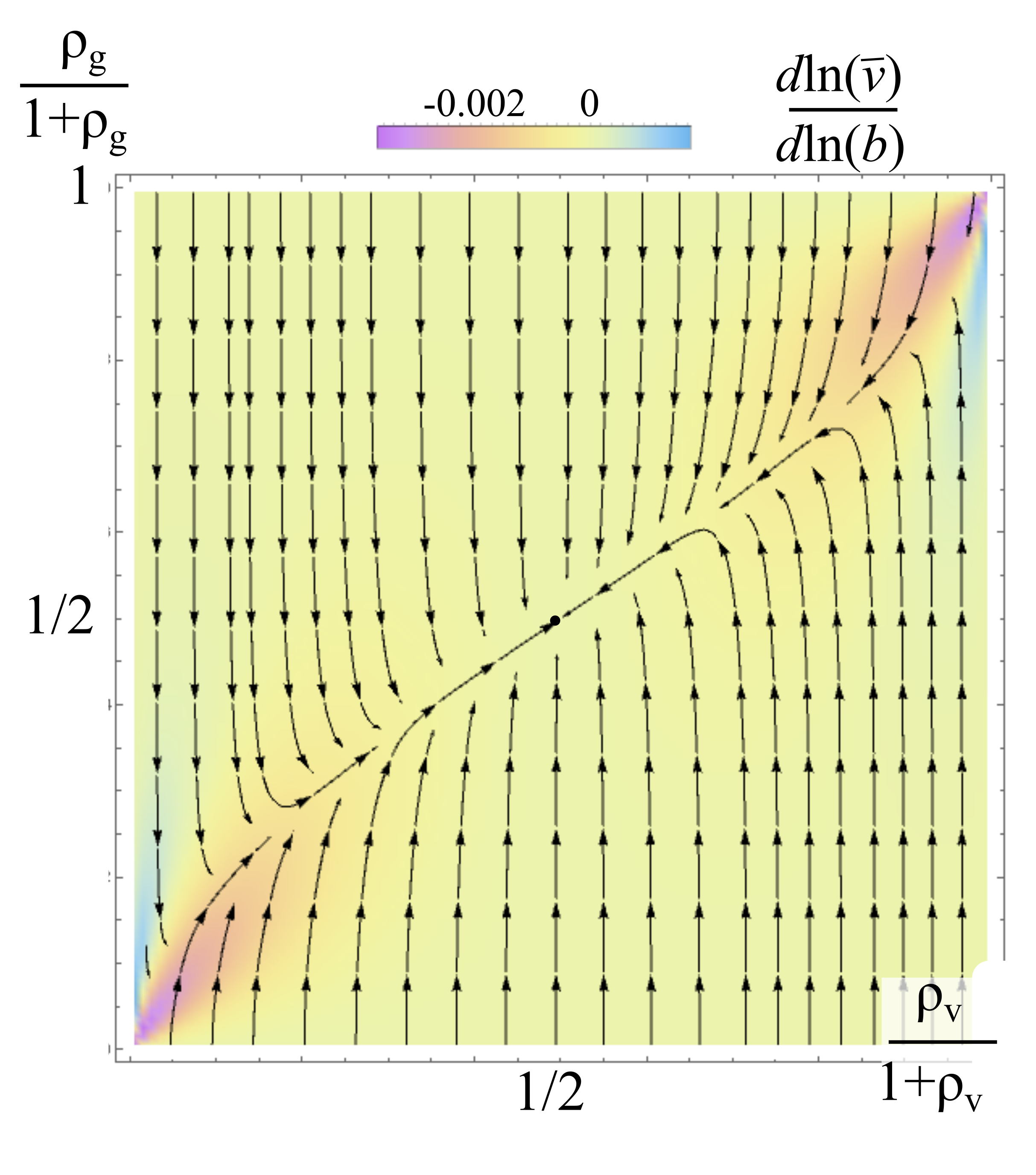}
    \caption{RG flow, Eqs.~\eqref{eq:BetaRhog}, \eqref{eq:BetaRhov} in the plane spanned by $\rho_v, \rho_g$ in the case $N_+ = N_-$. The color code illustrates the strength of the renormalization of the average velocity $\bar{v}$.} 
    \label{fig:RGFlowEqualN}
\end{figure}

\subsection{RG flow in symmetric case}

In the main text we discussed the RG flow for the case $N_+ \neq N_-$ which is most relevant to twisted trilayer graphene. Here, we briefly discuss the case $N_+ = N_-$, see Fig.~\ref{fig:RGFlowEqualN}. Note that in contrast to Fig.~\ref{fig:RGFlow} of the main text, the RG flow is now symmetric under $\rho_{v,g} \leftrightarrow 1/\rho_{v,g}$.

\subsection{RG flow near $\rho_v = 1$}

The RG equations can be calculated exactly to leading order in $\rho_v = 1 + \delta \rho_v$, in which case
\begin{align}
    \frac{d \delta \rho_v}{d \ln(b)} &= -\frac{32 \rho_g^2}{15N \pi ^2 \left[n \left(\rho_g^2-1\right)+\rho_g^2+1\right]^2}  \delta \rho_v,\\
    \frac{d \rho_g}{d \ln(b)} &= -\frac{16 \rho_g \left(\rho_g^2-1\right)}{3N \pi ^2 \left[n \left(\rho_g^2-1\right)+\rho_g^2+1\right]} \notag \\
    &+\frac{64  \rho_g^3}{5 N \pi ^2 \left [n \left(\rho_g^2-1\right)+\rho_g^2+1\right]^2} \delta \rho_v ,\\
    \frac{d \ln(\bar v)}{ d\ln(b)} &= \frac{16  n \rho_g^2}{15N \pi ^2 \left[n \left(\rho_g^2-1\right)+\rho_g^2+1\right]^2}\delta \rho_v,
\end{align}
where $n = (N_+ - N_-)/(N_+ + N_-)$. Near the isotropic point, we can expand these RG equations for $\delta \vec \rho= (\delta \rho_v, \delta \rho_g)$, both $\delta \rho_{v,g} = \rho_{v,g} - 1$,
\begin{align}
    \frac{d \delta \vec \rho}{d \ln(b)} &= - \frac{8}{15 \pi^2 N}\left (\begin{array}{cc}
        1 & 0 \\
        -6 & 10
    \end{array}\right) \delta \vec \rho.
\end{align}
This leads to scaling dimensions $- \frac{8}{15 \pi^2 N}$ [$-\frac{80}{15 \pi^2 N}$] with eigenvectors $(3,2)/\sqrt{13}$ [(0,1)], independently on $n$.

\end{document}